\documentclass[baaa]{baaa}

\usepackage[pdftex]{hyperref}
\usepackage{subfigure}
\usepackage{natbib}
\usepackage{helvet,soul}
\usepackage[font=small]{caption}

\contriblanguage{1}

\contribtype{1}

\thematicarea{1}

\title{The Fornax Cluster through S-PLUS}

\titlerunning{The Fornax Cluster through S-PLUS}

\author{A.V. Smith Castelli\inst{1,2}, C. Mendes de Oliveira\inst{3}, F. Herpich\inst{3}, C.E. Barbosa\inst{3}, C. Escudero\inst{1,2},  M. Grossi\inst{4}, L. Sodr\'e\inst{3}, C.R. de Bom\inst{5}, L. Zenocratti\inst{1,2}, M.E. De Rossi\inst{6}, A. Cortesi\inst{4},  R. Cid Fernandes\inst{7}, A.R. Lopes\inst{8},  E. Telles\inst{8}, G.B. Oliveira Schwarz\inst{9}, M.L.L. Dantas\inst{10}, F.R. Faifer\inst{1,2}, A. Chies Santos\inst{11}, J. Saponara\inst{12}, V. Reynaldi\inst{1}, I. Andruchow\inst{1,2}, L. Sesto\inst{1,2}, M.F. Mestre\inst{1,2}, A.L. de Amorim\inst{7}, E.V.R. de Lima\inst{3}, J.C.R. Abboud\inst{3}, V. Cernic\inst{3} \& I. Souza de Almeida Garcia\inst{3}}

\contact{asmith@fcaglp.unlp.edu.ar}

\institute{
Facultad de Ciencias Astron\'omicas y Geof\'isicas, UNLP, Argentina \and
Instituto de Astrof\'isica de La Plata, CONICET--UNLP, Argentina \and
Instituto de Astronomia, Geof\'isica e Ci\^encias Atmosf\'ericas, USP, Brasil \and
Observatorio do Valongo, UFRJ, Brasil \and
Centro Brasileiro de Pesquisas F\'isicas, Brasil \and
Insituto de Astronom\'ia y F\'isica del Espacio, CONICET--UBA, Argentina \and
Departamento de F\'isica, UFSC, Brasil \and
Observatorio Nacional, Brasil \and
Universidade Anhembi Morumbi, Brasil \and
Nicolaus Copernicus Astronomical Center, Polish Academy of Sciences, Polonia \and
Departamento de Astronom\'ia, UFRGS, Brasil \and
Instituto Argentino de Radioastronom\'ia, CONICET--CICPBA--UNLP, Argentina 
}

\resumen{El relevamiento S-PLUS (Southern Photometric Local Universe Survey) tiene como objetivo mapear $\approx 9300~\text{deg}^2$ del Hemisferio Sur Celeste usando el sistema de filtros Javalambre. Dicho sistema consiste de 12 bandas \'opticas, 5 de las cuales son similares a las utilizadas por el Sloan Digital Sky Survey (SDSS), mientras que las 7 restantes corresponden a filtros de banda angosta que trazan zonas específicas del espectro óptico ([OII], Ca H+K, D4000, H$\delta$, Mgb, H$\alpha$ and CaT). S-PLUS es llevado adelante con el telescopio robótico T80-South de 0.826-m, el cual se encuentra alojado en CTIO y est\'a equipado con una c\'amara de campo amplio (FoV $\approx2~\text{deg}^2$). En este p\'oster presentamos el projecto \#59 de la colaboraci\'on S-PLUS, cuyo objetivo es el de estudiar el c\'umulo de galaxias de Fornax, cubriendo un \'area del cielo equivalente a $\approx 11 \times 7 ~\text{grados}^2$, y con fotometr\'ia homog\'enea en las 12 bandas \'opticas de S-PLUS (Coordinadora: A. Smith Castelli). 
}

\abstract{
The Southern Photometric Local Universe Survey (S-PLUS) aims to map $\approx 9300~\text{deg}^2$ of the Southern sky using the Javalambre filter system of 12 optical bands, 5 Sloan-like filters and 7 narrow-band filters centered on several prominent stellar features ([OII], Ca H+K, D4000, H$\delta$, Mgb, H$\alpha$ and CaT). S-PLUS is carried out with the T80-South, a new robotic 0.826-m telescope located on CTIO, equipped with a wide FoV camera (2 deg$^2$). In this poster we introduce project \#59 of the S-PLUS collaboration aimed at studying the Fornax galaxy cluster covering an sky area of $\approx 11 \times 7~\text{deg}^2$, and with homogeneous photometry in the 12 optical bands of S-PLUS (Coordinator: A. Smith Castelli). 
}


\begin{document}

\maketitle

\section{Introduction}
\label{S_intro}
\subsection{The Fornax cluster}
The Fornax galaxy cluster ($(m-M)=31.51$; \citealt{Blakeslee2009}) is the second richest association within 20 Mpc after Virgo. It consists of a main structure centered in NGC\,1399 ($<V_r>=1442~\text{km}~ \text{s}^{-1}$), and an infalling substructure around NGC\,1316 (Fornax\,A) ($<V_r>=1778~\text{km}~\text{s}^{-1}$) \citep{Maddox2019}. According to \citet{Zabel2020}, {\it "... Fornax is the smaller sibling of the Virgo cluster"} as it is located at a similar distance and it has 1/10 of its mass ($\approx7 \times 10^{13}~\text{M}_\odot$). Despite Fornax contains 1/6 of the amount of galaxies found in Virgo, it is 2-3 times denser, more symmetric and more dinamically evolved (\citealp{Zabel2020} and references therein). All these characteristics make Fornax an interesting environment in which galaxy formation and evolution scenarios can be tested in the southern sky. 

\subsection{S-PLUS}

The Southern Photometric Local Universe Survey (S-PLUS) is a joint scientific effort of Brazilian, Chilean and Spanish institutions that aims to map $\approx 8000~\text{deg}^2$ of the Southern sky with twelve optical filters consisting of 5 broadband SDSS filters and 7 narrow-band filters centered on several prominent features (i.e., [OII], Ca H+K, D4000, H$\delta$, Mgb, H$\alpha$ and CaT) \citep{Claudia2019}. The multi-purpose astrophysical survey in the southern hemisphere has started at the end of 2016. S-PLUS data are ideal for searching low-metallicity and blue-horizontal branch stars and high redshift quasars, study the star formation histories of large numbers of galaxies selected based on the accurate 12-band photometric redshifts, and to map the large scale structure in the nearby universe, among other science goals. 

S-PLUS is carried out with the T80-South, a new 0.826 m telescope optimised for robotic operation, equipped with a wide FoV camera (2 deg$^2$; 0.55 arcsec pixel$^{-1}$). The telescope, camera, and filter set are identical to those of the Javalambre Auxiliary Survey Telescope (T80/JAST), installed at the Observatorio Astrofísico de Javalambre. T80/JAST is currently performing the Javalambre Photometric Local Universe Survey (J-PLUS), a 12-band survey of a complementary area in the northern hemisphere.

The first data release (DR1) of S-PLUS is already available and contains 80 Stripe-82 fields observed during the scientific validation process plus other 90 fields of the same region obtained during regular operation time\footnote{\url{http://www.splus.iag.usp.br/data/}}. Other releases adding entire regions of the southern sky will be followed in the upcoming months.

\section{The S-PLUS Fornax Project}

\begin{figure}[!t]
\centering
\includegraphics[width=\columnwidth]{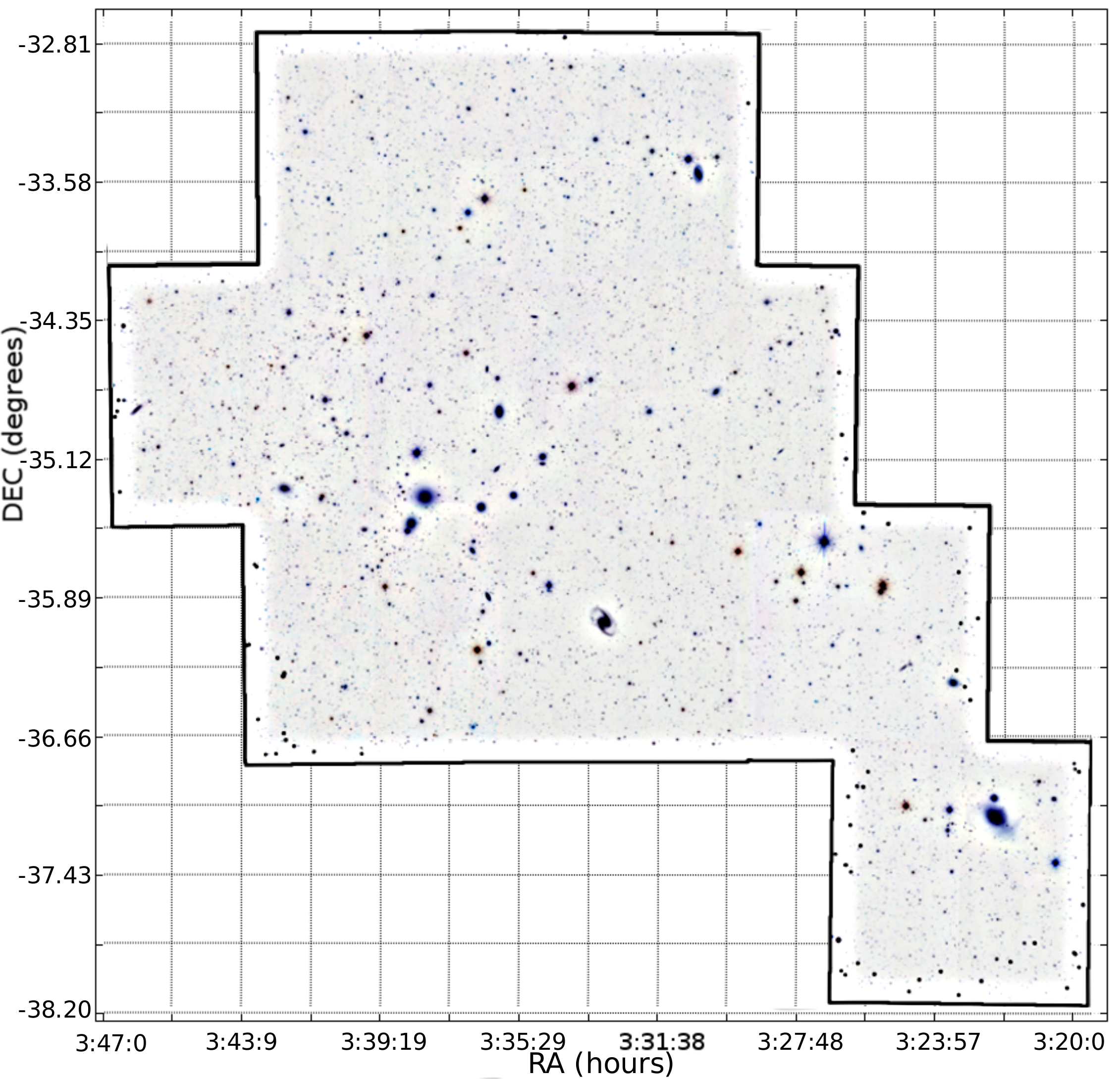}
\caption{Mosaic of the 9 S-PLUS fields covering the two main structures of the Fornax cluster. These 9 fields equals a sky area of $6.75 \times~5.5~\text{deg}^2$. North is up and East is to the left.}
\label{Fornax_fields}
\end{figure}

The data of our project consist on 23 S-PLUS fields observed in the 12 photometric bands of the survey. The central main structure of Fornax and the infalling subgroup of NGC\,1316 are covered by 9 of those fields ($6.75 \times 5.5~\text{deg}^2$;  Figure\,\ref{Fornax_fields}). The additional 14 S-PLUS fields correspond to the outskirts of the cluster, and they will help to reach more distant Fornax galaxies included in several catalogues reported in the literature (e.g. \citealt{Ferguson1989,Venhola2018,Maddox2019}). We plan to complement these data with UV, NIR and radio observations.  

\begin{figure}[!t]
\centering
\includegraphics[width=\columnwidth]{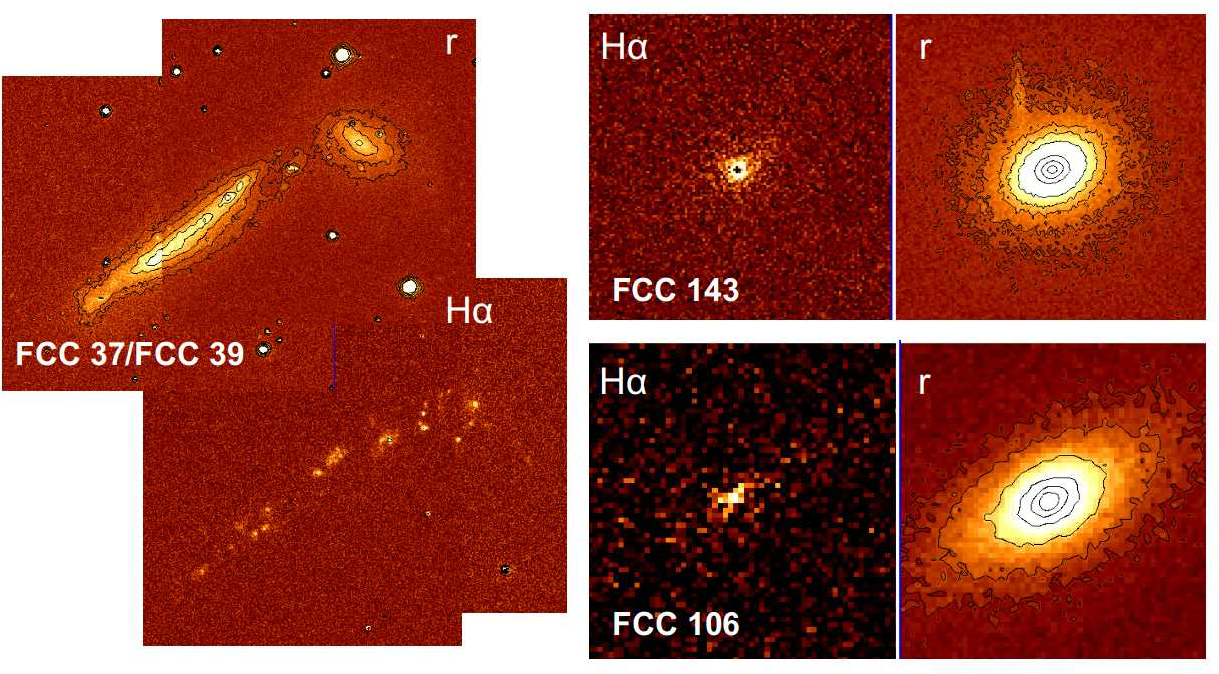}
\caption{S-PLUS images of different classes of dwarf galaxies in Fornax. {\it Left panels:} dwarf irregulars FCC37/FCC39. {\it Right panels:} dwarf ellipticals FCC106 and FCC143. For each galaxy, we show the r-band and the continuum-subtracted H$\alpha$ images.}
\label{Dwarfs_marco}
\end{figure}

The topics to be covered in the framework of our Fornax project include, among others, the identification of H$\alpha$ emitters, low surface brightness galaxies, bright and compact objects, new dwarf and H\,II galaxies and peculiar systems. In particular, as dwarf galaxies (Figure\,\ref{Dwarfs_marco}) are sensitive probes of the environment where they are evolving due to the fragility of their stellar structures, we expect that their study in a rich cluster such as Fornax will contribute to understand whether the interaction between galaxies and the intergalactic medium (ram pressure stripping) is dominant compared to galaxy-galaxy interactions (tidal interactions, harassment). Furthermore, comparing Fornax with clusters in different evolutionary stages will provide a more complete view of the role of these different processes according to the cluster properties.

\begin{figure}[!t]
\centering
\includegraphics[width=\columnwidth]{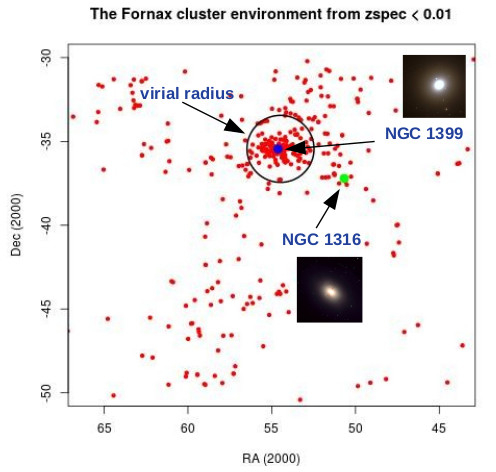}
\caption{Large scale structure around the Fornax cluster within $0 < z <0.01$, showing galaxies with spectroscopic redshift from NED. The blue point corresponds to the bright cluster galaxy NGC\,1399 and the green dot indicates the position of NGC\,1316 (Fornax\,A). The circle depict the virial radius of Fornax. S-PLUS will allow to investigate the galaxy evolution along the filaments that feed the
cluster.}
\label{LSS_laerte}
\end{figure}

\begin{figure}[!t]
\centering
\includegraphics[width=\columnwidth]{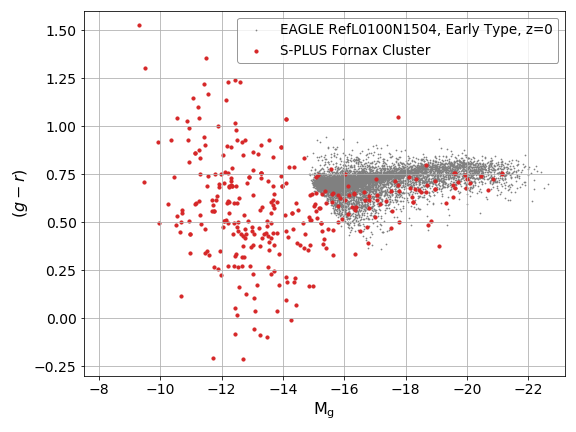}
\caption{Comparison between the color-magnitude relation defined by the early-type galaxies in Fornax, and the red sequence followed by simulated early-type galaxies selected in the EAGLE simulation.}
\label{Simus_Lucas}
\end{figure}

Using radial velocity catalogues available in the literature, and thanks to the wide FOV of S-PLUS and the coverage with several pointings around Fornax, it will be possible to explore the filaments that feed the cluster (Figure\,\ref{LSS_laerte}) as well as to analyze the dynamical properties of the different types of galaxies. In addition, through the identification of Fornax-like galaxies in state-of-the-art numerical simulations (Figure\,\ref{Simus_Lucas}), we will follow their formation histories and propose formation scenarios for specific morphological types of galaxies in dense environments. We will also select galaxy clusters with the main characteristics of Fornax to analyze its possible evolutionary path.

\begin{figure}[!t]
\centering
\includegraphics[width=0.48\columnwidth]{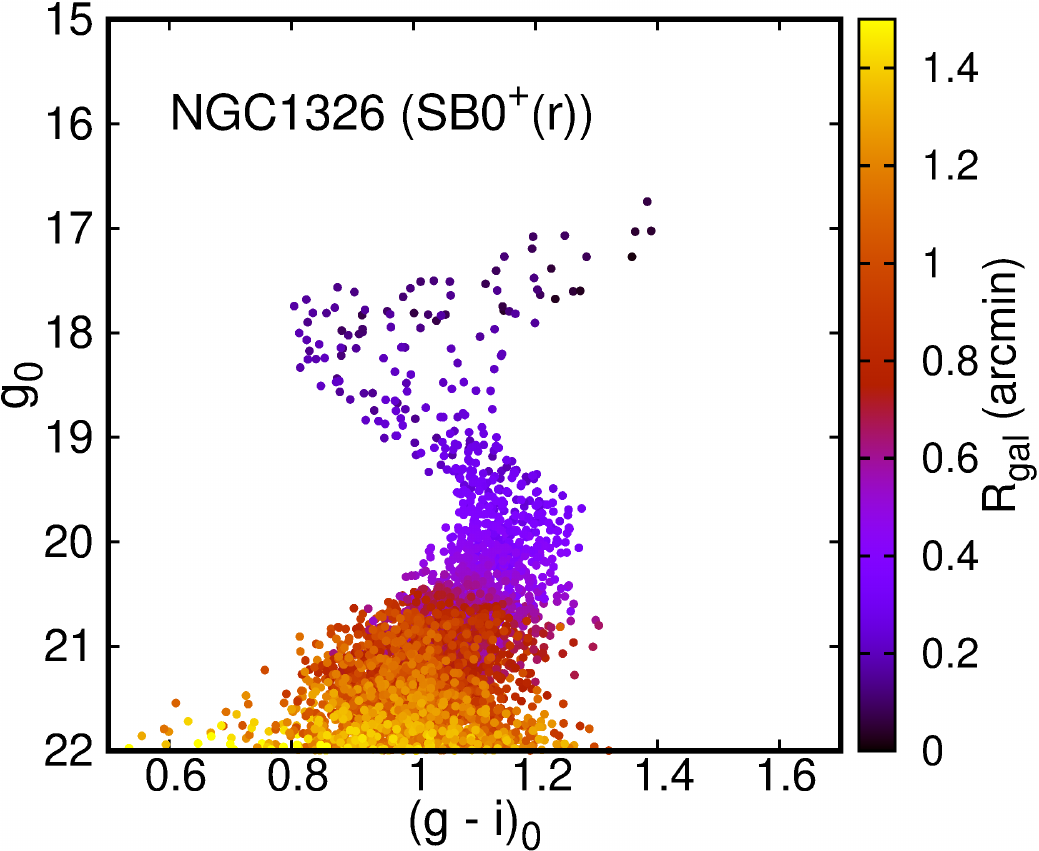}
\includegraphics[width=0.48\columnwidth]{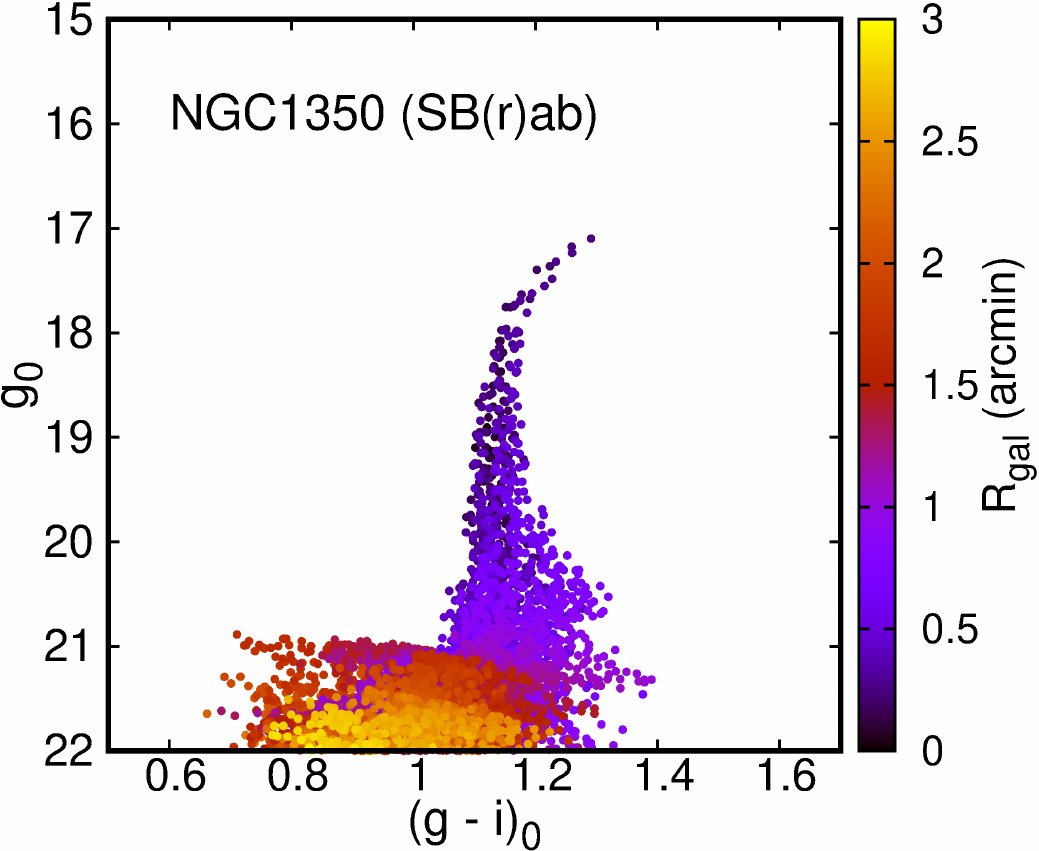}
\includegraphics[width=0.48\columnwidth]{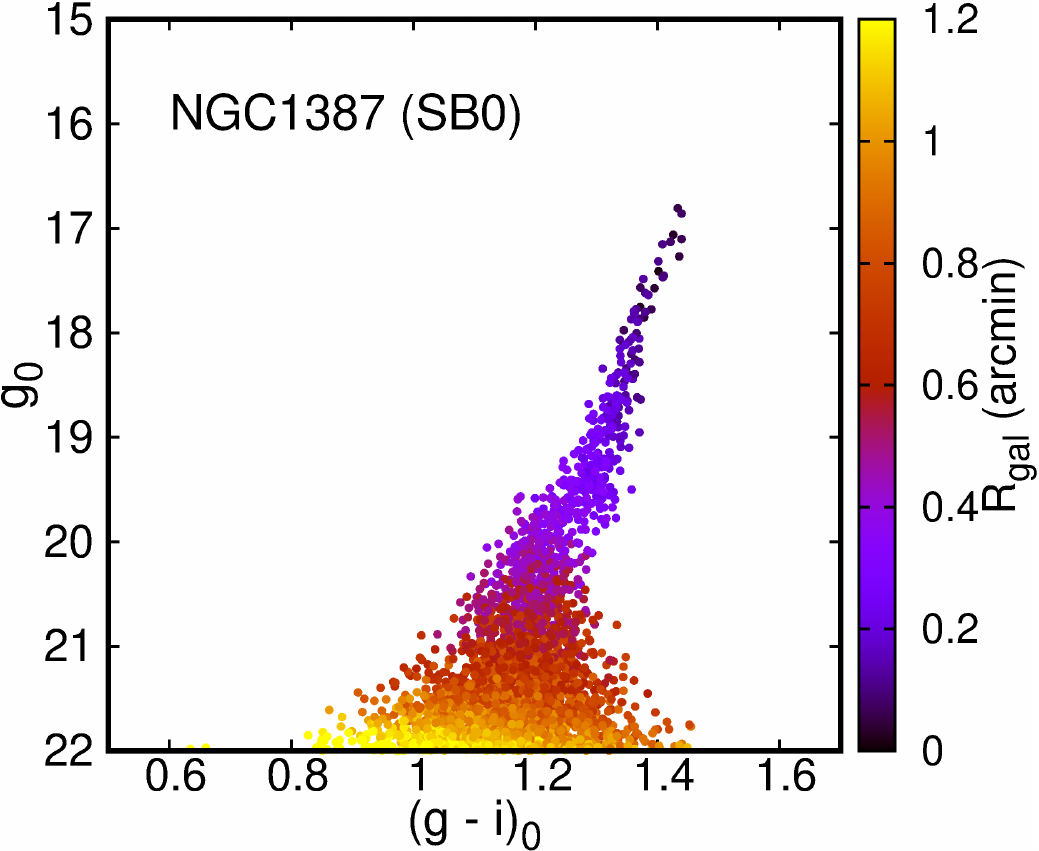}
\includegraphics[width=0.48\columnwidth]{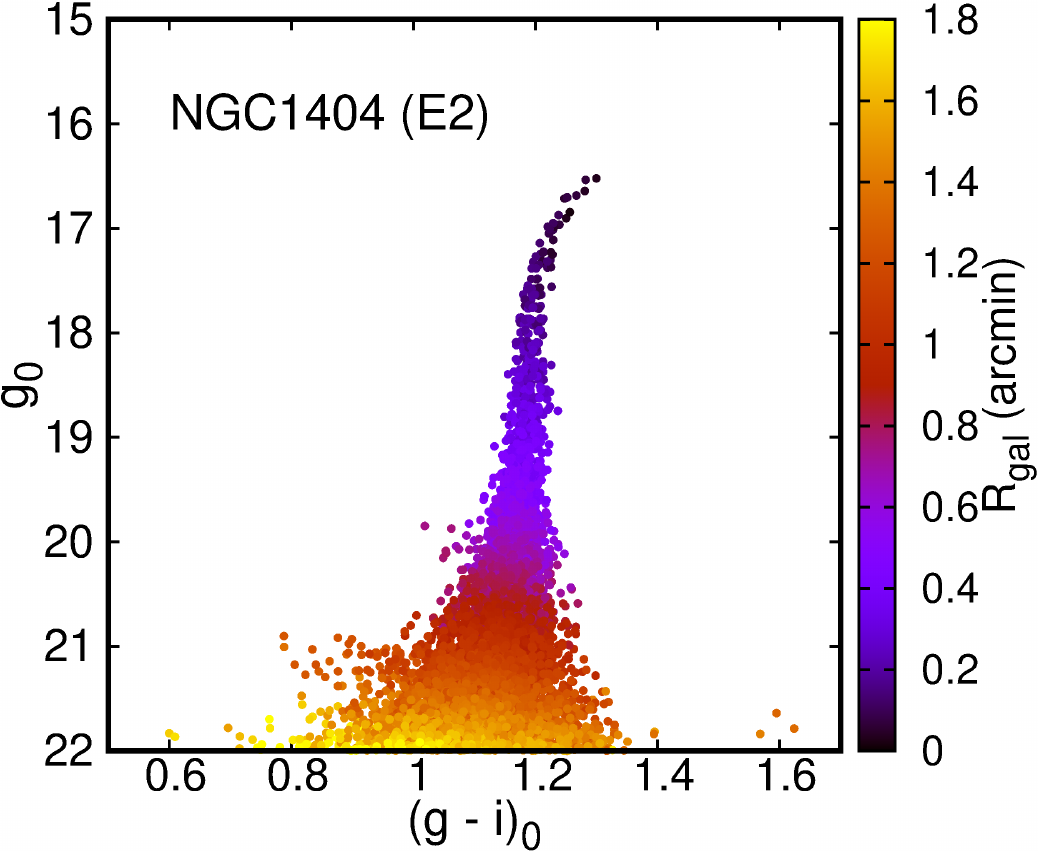}
\caption{(g-i) vs. g pixel color-magnitude diagrams obtained from S-PLUS images of four galaxies of different morphological types. The right color bars indicate the radial distance of each pixel to the center of the galaxy. From left to right, and from top to bottom, we show the diagrams of two ring galaxies (NGC\,1326 and NGC\,1350), a barred lenticular galaxy (NGC\,1387) and an elliptical galaxy (NGC\,1404).}
\label{PxCMD}
\end{figure}

From a technical point of view, we will also explore the application of different analysis techniques, such as pixel color-magnitude diagrams (\citealp{Lee2017}; Figure\,\ref{PxCMD}) to try to unveil the inner structures of the brightest galaxies. In addition, due to the large amount of images that we have to handle, we will use specific software to semi-automatically obtain structural parameters, such as {\rm GALAPAGOS-C} \citep{Hiemer2014}. This kind of techniques might help to perform a detailed morphological classification of the galaxies in the sky region of Fornax. 

\section{Impact and Future Perspectives}
Though highly studied, Fornax has never been observed with the combination of a wide field of view ($1.4 \times 1.4~\text{deg}^2$) and simultaneous 12 photometric bands. The large sky coverage achieved with S-PLUS will allow us to explore not only the main substructures of the cluster but also its outskirts with great detail. Furthermore, the wide FOV of the images obtained by S-PLUS impose reduction and data-handle challenges that might promote the development and/or application of novel techniques, such as deep and machine learning. We expect this whole analysis will provide information to better understand the dynamics of the Fornax cluster, and the relation between different types of galaxies and the cluster environment. 

\begin{acknowledgement}
We would like to thank the referee for her/his comments about the manuscript. 
S-PLUS is an international collaboration founded by Universidade de Sao Paulo, Observat\'orio Nacional, Universidade Federal de Sergipe, Universidad de La Serena and Universidade Federal de Santa Catarina.  
This work was funded with grants from Consejo Nacional de Investigaciones Cient\'ificas y T\'ecnicas de la Rep\'ublica Argentina, and Universidad Nacional de La Plata (Argentina).
\end{acknowledgement}

\bibliographystyle{baaa}
\small
\bibliography{565_v3}
 
\end{document}